\documentstyle [preprint,aps,eqsecnum]{revtex}

\begin{document}

\preprint{UW/PT-92-23}

\title{$2 \to n$ threshold production at tree level}

\author{Lowell S. Brown and Chengxing Zhai}

\address{ Department of Physics, FM-15,
    University of Washington,
    Seattle, Washington 98195
    }%
\maketitle

\begin{abstract}
The threshold behavior of a theory with two coupled scalar fields $\chi$ and
$\phi$ is investigated. We compute the amplitude for two on-mass-shell
$\chi$-particles to produce an arbitrary number of $\phi$-particles at
rest in the tree approximation.
\end{abstract}
\newpage
\section{Introduction}
The solution of the classical field equation in the presence of a source
yields the generating functional for tree graphs. This feature of
quantum field theory has recently been exploited to provide a very simple
derivation \cite{Brown} of the tree-level $1 \to n$ threshold production
amplitudes, $\langle n | \phi | 0 \rangle$, that had been previously
obtained by other means \cite{Voloshin0,Argyres}.
This method has been extended \cite{Voloshin1,Voloshin2,Voloshin3,Smith} to
calculate the
$2 \to n$ threshold production amplitude at the tree level for the
$\lambda \phi^4$ theory by using this classical solution as
a background field.
The interesting result is that for $\lambda \phi^4$ theory with unbroken
$\phi \to -\phi$ symmetry, the tree-level amplitude for $2 \to n$ threshold
production vanishes for $n > 4$. For the spontaneous
symmetry breaking case, this amplitude vanishes for $n > 2$.
These null amplitude results depend upon the fact that in the pure
$\lambda \phi^4$ theory a dimensionless factor takes on a special value.

In this article, we shall consider a more general case of the $2 \to n$
process in which the incoming particle is described by a $\chi$-field
that is not required to be the same as the final produced particles
described by the $\phi$-field.\cite{Argyres2}
Thus we shall study a coupled field system
containing more parameters than in the ordinary $\lambda \phi^4$ theory,
which enables the dimensionless factor mentioned above to take on
arbitrary values. The amplitude for $2 \to n$ threshold production is,
in general, not zero.
This nonvanishing amplitude will be calculated explicitly, from which
we will see that the dimensionless parameter does play
a critical role in the vanishing of the amplitude for the pure $\lambda
\phi^4$ theory.
Our method to compute the $2 \to n$ threshold production amplitude is to
first use the reduction formula to express the amplitude
in terms of the Fourier transform of the two-point Green's
function of the incoming $\chi$-field, with
the classical solution of the $\phi$-field appearing as a
background field. This Green's function
can be regarded as the inverse of a differential operator
which defines a quantum-mechanical scattering problem.
Thus the desired Fourier transform of the Green's function is related
to a scattering amplitude which can be obtained from the solution of the
homogeneous Klein-Gordon equation. This solution is easily obtained in
terms of a hypergeometric function, and identifying the corresponding
scattering amplitude gives the $2 \to n$ threshold production amplitude
at the tree-level in terms of ratios of gamma functions. Our method
simplifies and generalizes previous work
\cite{Voloshin1,Voloshin2,Voloshin3,Smith}, and along the way clarifies
the technique introduced in \cite{Brown}.

\section{General Formulation}
Previous work dealt with the threshold production in tree
approximation of many scalar particles for the theory
described by the Lagrange function
\begin{equation}
    {\cal L}_{\phi} = - {1 \over 2} (\partial \phi)^2
	- {1 \over 2} m^2 \phi^2 - {1 \over 4!} \lambda \phi^4
	+ \rho \phi \,.
\end{equation}
An external source $\rho(x)$ is introduced in order to generate the
multiparticle amplitudes. Our purpose here is to generalize this previous
work by introducing an additional scalar field $\chi$ of mass $M$ which is
coupled to the scalar field $\phi$. We shall compute the amplitude
for two high-energy $\chi$-particles (on their mass shell) to produce many
on-mass-shell $\phi$-particles at rest. We take the Lagrange function for
the coupled field system to be given by
\begin{equation}
    {\cal L} = - {1 \over 2} (\partial \chi)^2
	- {1 \over 2} M^2 \chi^2 - {1 \over 6} a(a-1){1 \over 4}
	\lambda \chi^2 \phi^2 + {\cal L}_{\phi} \,.
\end{equation}
As will be seen, it proves convenient to write the $\chi^2 \phi^2$
coupling parameter in
the way in which we have done here. In particular, when $a=3$, the
$\chi$-field describes the small fluctuations about the classical field
$\phi$ produced by the source $\rho$, and in this limit the amplitude
which we shall compute reduces to the $2 \phi \to$ many $\phi$ threshold
amplitude in tree approximation.

The $2 \chi \to$ many $\phi$ amplitude in tree approximation may be
obtained from the generating functional
\begin{eqnarray}
    \langle 0 + | p \, p^{\prime} - \rangle^{\rho} &=&
	i(p^2 + M^2) \int (d^4x) \, e^{i px} \,
	 (p^{\prime 2} + M^2) \int (d^4x^{\prime})
	\, e^{i p^{\prime}x^{\prime}}
\nonumber\\
	&& \times \,\langle 0 + |
	i {\cal T} \chi (x) \chi (x^{\prime})
	| 0 - \rangle^{\phi_{\rm cl}} \,.
\label{reduct}
\end{eqnarray}
Here the reduction method has been used to construct the initial two
$\chi$-particle state with the four-momenta $p$ and $p^{\prime}$.
The mass-shell limits $-p^2 \to M^2$ and $ - p^{\prime2} \to M^2$
are to be taken once the corresponding poles in the Green's function
\begin{equation}
    G (x, x^{\prime}) = \langle 0 + |
	i {\cal T} \chi (x) \chi (x^{\prime})
	| 0 - \rangle ^{\phi_{\rm cl}}
\label{green}
\end{equation}
have been displayed. In the tree approximation, the source $\rho$
produces a classical field $\phi_{\rm cl}$ under which the $\chi$-field
propagates. A final application of the reduction method yields
the $n$ $\phi$-particle threshold production amplitude $A_n$,
\begin{eqnarray}
    i (2 \pi)^4 \delta (nm \,&\!&- E - E^{\prime}) \,
	\delta ({\bf p} + {\bf p}^{\prime}) \, A_n
\nonumber\\
	= &\!&\left . \prod_{a=1}^n \int (d^4 x_a) e^{i \omega t_a}
	(m^2 - \omega^2) {\delta \over \delta \rho(x_a)}
	\langle 0 + | p \, p^{\prime} - \rangle ^{\rho}
	\right |_{\rho = 0} \,,
\end{eqnarray}
since each functional derivative with respect to the source $\rho (x)$ inserts
a time-ordered factor of $i\phi (x)$ into the matrix element. After the
appropriate poles have been identified, the mass-shell limit $\omega
\to m$ is to be taken.

The construction
\begin{equation}
    {\cal A} (z_0) = \left . \exp \left \{z_0 \int (d^4 x) e^{i \omega t}
	(m^2 - \omega^2) {\delta \over \delta \rho (x)} \right \}
	\langle 0 + | p \, p^{\prime} - \rangle ^{\rho}
	\right |_{\rho = 0}
\label{trick}
\end{equation}
generates the threshold amplitudes by ordinary differentiation
with respect to the parameter $z_0$.
The exponential operation which appears here is just the functional
Taylor series expansion  which has the effect of the source replacement
\begin{equation}
    \rho (x) \to \rho_0 (t) = z_0 (m^2 - \omega^2) e^{i \omega t} \,.
\label{replace}
\end{equation}
Thus the generating function (not functional) ${\cal A} (z_0)$
is given by making this source replacement in the original generating
functional,
\begin{equation}
    {\cal A} (z_0) =
        \langle + 0 | p \, p^{\prime} - \rangle ^{\rho_0} \,.
\end{equation}

In the tree approximation,
a source driven amplitude is a functional of the classical field,
and thus
\begin{equation}
    \langle 0 + | p \, p^{\prime} - \rangle^{\rho}
	= {\cal F} [\phi_{\rm cl}] \,,
\end{equation}
where
\begin{equation}
    ( - \partial ^2 + m^2 ) \phi_{\rm cl}
	+ {1 \over 3!} \lambda \phi_{\rm cl}^3 = \rho \,.
\label{classeq}
\end{equation}
To see the consequence of the source replacement~(\ref{replace}),
we note that in the $\lambda = 0$, free-field limit, the solution
of the classical field equation~(\ref{classeq}) becomes
\begin{equation}
    \phi_{\rm cl} \to z(t) = z_0 e^{i \omega t} \,.
\label{limit}
\end{equation}
In view of the structure of the classical field equation~(\ref{classeq}),
the high-order terms in the coupling expansion of the field $\phi_{\rm cl}$
are driven by $z(t)$.  Thus, the mass-shell limit $\omega \to m$ may now be
taken with $\phi_{\rm cl}$ obeying the homogeneous, ordinary differential
equation
\begin{equation}
    \left [ {d^2 \over dt^2} + m^2 \right ]
	\phi_{\rm cl} (t) + {1 \over 3!} \lambda \phi_{\rm cl}^3 (t) = 0 \,,
\label{diffeq}
\end{equation}
subject to the conditions that $\phi_{\rm cl} (t)$ reduce to $z(t)$ when
$\lambda = 0$ and that the solution be analytic in $z(t)$ for small $z(t)$.

To obtain the many-particle amplitude with two initial $\chi$-particles, we
note that the Green's function defined by Eq.~(\ref{green}) obeys, in a
compact, operator-like notation,
\begin{equation}
    \left [ G_0 ^{-1} -  V \right ] G = 1 \,,
\label{Green}
\end{equation}
where
\begin{equation}
    G_0^{-1} = - \partial^2 + m^2  \,,
\end{equation}
and
\begin{equation}
    -  V = {1 \over 6} a(a-1){1 \over 2} \lambda
	\phi_{\rm cl}^2 (t) \,.
\end{equation}
The Green's function equation~(\ref{Green}) has the formal solution
\begin{equation}
    G = {1 \over 1 - G_0 V} G_0 \,,
\end{equation}
which, by simple operator algebra, may be arranged to read as
\begin{equation}
    G = G_0 + G_0 T G_0 \,,
\label{decomp}
\end{equation}
where
\begin{equation}
    T = V {1 \over 1 - G_0 V} \,.
\end{equation}
The first term on the right-hand side of Eq.~(\ref{decomp}) describes the
free propagation of the $\chi$-field, and it may be omitted. When Fourier
transformed, the $G_0$ factors which flank $T$ in the second term are simple
poles that cancel the $(p^2 + M^2)$ and $(p^{\prime 2} + M^2)$ factors in
Eq.~(\ref{reduct}). Thus we may now go on mass shell and write
\begin{eqnarray}
    {\cal A} (z_0) = \langle 0+ | p \, p^{\prime} - \rangle ^{\rho_0}
	&=& i \int (d^4x) \int (d^4x^{\prime}) e^{i p^{\prime} x^{\prime}}
	T (x^{\prime}, x) e^{i p x}
\nonumber\\
	&=& i T (-p^{\prime}, p) \,.
\label{scattered}
\end{eqnarray}
Expanding this amplitude in powers of $z_0$ yields the desired
multiparticle production amplitudes:
\begin{equation}
     {\cal A}(z_0) = \sum_{n=0}^{\infty} i \, (2 \pi)^4
        \delta (nm - E - E^{\prime}) \, \delta ({\bf p} +
        {\bf p}^{\prime}) A_n \, {z_0 ^n \over n!} \,.
\label{generatfun}
\end{equation}

As is usual in quantum-mechanical scattering theory, it is easier to extract
the ``scattering amplitude'' $T$ from the solution $\psi$ of the homogeneous
wave equation
\begin{equation}
    \left [ G_0 ^{-1} - V \right ] \psi = 0 \,,
\label{waveeqn}
\end{equation}
rather than dealing directly with the Green's function $G$. The wave
equation may be written as an integral equation,
\begin{equation}
    \psi = \psi_0 + G_0 V \psi \,,
\end{equation}
with $\psi_0$ being a free solution,
\begin{equation}
    G_0^{-1} \psi_0 = 0 \,.
\end{equation}
Iteration of the integral equation produces the formal solution
\begin{eqnarray}
    \psi &=& \psi_0 + G_0 V {1 \over 1 - G_0 V} \psi_0
\nonumber\\
	&=& \psi_0 + G_0 T \psi_0 \,.
\label{result}
\end{eqnarray}
Thus we solve the wave equation~(\ref{waveeqn})
subject to the conditions that $\psi$
reduces to $\psi_0 = e^{ipx}$ in the absence of interaction and that
$\psi$ be analytic in $z(t)$ for small $z(t)$. Then, in view of the
structure of the formal solution~(\ref{result}), $\psi$ will have
a pole corresponding to $-p^{\prime 2} = M^2$ as shown by the $G_0$
factor on the right-hand side of Eq.~(\ref{result}), and the residue
of this pole will define the amplitude $T(-p^{\prime}, p)$.
The details of this procedure will be made clear in the examples which
we now turn to work out.

\section{Unbroken Symmetry}
With $m^2 > 0$, the reflection symmetry $\phi \to - \phi$ is not broken,
and the proper solution of the differential equation~(\ref{diffeq})
for the classical field is given by \cite{Brown}
\begin{equation}
    \phi_{\rm cl} (t) = { z(t) \over 1 - (\lambda / 48 m^2) z(t)^2} \,,
\end{equation}
where
\begin{equation}
    z(t) = z_0 e^{i m t} \,.
\end{equation}
This classical field appears in a ``potential-like'' term in the
wave equation for $\psi$,
\begin{equation}
    \left [ - \partial^2 + M^2 + {1 \over 6} a(a-1)
	{1 \over 2} \lambda \phi_{\rm cl}^2 (t) \right ]
	 \psi = 0 \,.
\label{waveeq}
\end{equation}
This wave equation is spatially uniform and depends upon the
time only through the variable $z(t)$. Hence
its solution may be expressed as
\begin{equation}
    \psi (x) = e^{i px} F_1 (z (t)) \,.
\end{equation}
The conditions that $\psi \to \psi_0 = e^{ipx}$ in the absence of interaction
and that $\psi$ be analytic in $z(t)$ for small $z(t)$ require that
$F_1 (0) = 1$ and that $F_1 (z)$ has a convergent power series development
for small $z$.
It is, however, simpler to work with the new variable
\begin{equation}
    y(t) = - (\lambda / 48 m^2) z(t) \phi_{\rm cl} (t) \,,
\end{equation}
because of the relations
\begin{equation}
    {d \over dt} y(t) = 2 i m \, y(t) \left [ 1 - y(t) \right ] \,,
\end{equation}
and
\begin{equation}
    {1 \over 2} \lambda \phi_{\rm cl}^2 = - 24 m^2 \,
	y(t) \left [ 1 - y(t) \right ] \,.
\end{equation}
On writing
\begin{equation}
    \psi = e^{ipx} F_2 (y (t))
\label{magic}
\end{equation}
in the wave equation~(\ref{waveeq}) and using these relations,
a simple calculation yields
\begin{equation}
    y (1 -y ) F_2^{\prime \prime} (y)
	+ \left [1 - (E/m) - 2y \right ]
	F_2^{\prime}(y) - a(1-a) F_2 (y) = 0 \,,
\end{equation}
where the prime denotes the derivative.
This is the hypergeometric differential equation and thus the proper
solution is given by the hypergeometric function,
\begin{eqnarray}
    F_2 (y(t)) &=& _2 F_1 (a, 1 - a; 1 - E/m; y(t))
\nonumber\\
	&=& \sum_{k=0}^{\infty}
	{\Gamma (a+k) \over \Gamma (a)}
	{\Gamma (1-a+k) \over \Gamma (1-a)}
	{\Gamma(1 - E/m ) \over \Gamma (1 - E/m + k)}
	{y(t)^k \over k!} \,.
\label{hyper}
\end{eqnarray}

To extract the threshold production amplitude $A_n$ from this solution,
we recall from Eq.~(\ref{generatfun}) that the amplitude $A_n$ appears as a
coefficient of the term involving
\begin{equation}
    z(t)^n = z_0^n \, e^{i n m t} \,.
\end{equation}
Moreover, in view of the structure of Eq.~(\ref{result}), this term involves
the pole structure
\begin{eqnarray}
    G_0 \, e^{ipx} \, z(t)^n &=& {e^{ipx} z(t)^n \over
	E^2 - (nm - E)^2}
\nonumber\\
	&=& {e^{ipx} z(t)^n \over nm (2E - nm)} \,.
\end{eqnarray}
Finally, we note that the Fourier transform involving $p^{\prime}$
in Eq.~(\ref{scattered}) produces a factor of $(2\pi)^4 \delta (nm - E
- E^{\prime}) \delta ({\bf p} + {\bf p}^{\prime})$. From these remarks,
we conclude that $A_n/(nm\,n!)$ is the coefficient of the term which
contains $z(t)^n (2E-nm)^{-1}$ in the hypergeometric
function~(\ref{hyper}). To obtain this term, we note that
on the one hand,
\begin{equation}
    y(t)^k = (- \lambda/48m^2)^k z(t)^{2k} + \cdots \,,
\label{power}
\end{equation}
where the ellipsis stands for higher powers of $z(t)$, while, on the
other hand, the third gamma function ratio in the $k$-th term in the
hypergeometric function~(\ref{hyper}) may be written as
\begin{eqnarray}
    {\Gamma (1 - E/m) \over \Gamma(1 - E/m + k)}
	&=& {(-1)^k \over (k-1)!} \, {1 \over (E/m) - k} + \cdots
\nonumber\\
	&=& 2km \, {(-1)^k \over k!} \, {1 \over 2E - 2km} + \cdots \,,
\label{pole}
\end{eqnarray}
where the ellipses now stand for pole terms at lower energies,
$E = (2k - 2)m$, $E=(2k-4)m, \, \ldots \,$, and possible regular terms.
We find that the required pole structure\footnote{When one of the initial
$\chi$-particles is on mass shell but the other is off mass shell,
the tree structure of the final state has poles corresponding to tree
branches whose propagators go on the $\phi$-particle mass shell.
This accounts for the other poles which appear in the hypergeometric
function~(\ref{hyper}).}, namely a pole at $2E= nm$,
fits the required frequency dependence, $z(t)^n \sim \exp\{inmt\}$,
only for the leading terms displayed in
Eqs.~(\ref{pole})~and~(\ref{power})
with $n=2k$. The identification of the threshold production
amplitude $A_n$ in tree approximation is straightforward:
\begin{equation}
    A_{2k} = {\Gamma(a+k) \over \Gamma(a)} \,
	{\Gamma(1-a+k) \over \Gamma(1-a)} \,
	(2km)^2 \, (2k)! \left ({1 \over k!} \right )^2
	\left ({\lambda \over 48m^2} \right )^k \,.
\end{equation}

For general, non-integral, values of the coupling parameter $a$, the
threshold production amplitude for large number of particles,
$2k \gg 1$, has the factorial growth,
\begin{equation}
    A_{2k} \approx k \, (2k)! \, { (2 m)^2 \over \Gamma (a) \Gamma (1 - a)}
	\left ( {\lambda \over 48 m^2} \right )^k \,.
\end{equation}
However, if $a= N$ is an integer, the presence of the factor
\begin{equation}
    {\Gamma (1 - N + k) \over \Gamma (1 - N)} =
	(1-N)(2-N)\cdots (k-N)
\end{equation}
causes the amplitude for the production of $2k = 2N$ or more particles
at rest to vanish. In particular, the case where $a = 3$ gives the
amplitude for two initial on-mass-shell $\phi$-particles to produce $2k$
$\phi$-particles at rest. In this case, the threshold amplitude vanishes
except for the production of $2k=2$ or $2k=4$ particles. These results
on vanishing threshold amplitudes were previously found by
Voloshin \cite{Voloshin1,Voloshin2,Voloshin3}.

\section{Broken Symmetry}
Spontaneously symmetry breakdown occurs when the sign of the mass
term in the $\phi$-field Lagrange function is reversed, $m^2 \to -m^2$.
In this case, the solution to the classical field equation~(\ref{classeq})
describes the fluctuations about the shifted vacuum, constant field
solution of Eq.~(\ref{classeq}),
\begin{equation}
    \phi_{\rm cl} \to \phi_0 = \sqrt{{3! m^2 \over \lambda}} \,.
\end{equation}
This shift alters the mass in the field equation to $m_1 = \sqrt{2} m$,
and its proper solution now reads \cite{Brown}
\begin{equation}
    \phi_{\rm cl} (t) = \phi_0 + \sigma (t) \,,
\end{equation}
where
\begin{equation}
    \sigma (t) = {z(t) \over 1 - z(t) /2 \phi_0} \,,
\end{equation}
with
\begin{equation}
    z(t) = z_0 e^{i m_1 t} \,.
\end{equation}
The wave equation~(\ref{waveeq}) now becomes
\begin{equation}
    \left \{ - {\partial}^2 + M_1^2 + {1 \over 6}
	a(a-1) {1 \over 2} \lambda \left [
	\phi_{\rm cl} (t)^2 - \phi_0^2 \right ]
	\right \} \psi (x) = 0 \,,
\end{equation}
where $M_1^2 = M^2 + a(a-1)m^2/2$ denotes the new mass
for the $\chi$-field which has been shifted in the spontaneously
symmetry breakdown.
In parallel with our previous work, we write
\begin{equation}
    \psi (x) = e^{ipx} F_2 (y(t)) \,,
\end{equation}
where now
\begin{equation}
    y(t) = - \sigma (t) /2 \phi_0 \,.
\end{equation}
This is the appropriate variable because
\begin{equation}
    {d \over dt} y (t)  = i m_1 \, y(t) [1 - y(t)]
\end{equation}
and
\begin{equation}
    {1 \over 2} \lambda \left [ \phi_{\rm cl} (t)^2 - \phi_0^2
	\right ] = - 6 m_1^2 \, y(t) [1 - y(t)] \,.
\end{equation}
Thus we again find that $F_2 (y)$ obeys a hypergeometric
differential equation
\begin{equation}
    y (1 - y) F_2^{\prime \prime} (y)  + \left [ 1 - (2 E/ m_1) - 2y
	\right ] F_2^{\prime} (y) - a ( 1 - a) F_2 (y) = 0 \,.
\end{equation}
The appropriate solution is
\begin{equation}
    F_2 (y(t)) =\, _2 F_1 ( a, 1-a; 1 - 2E/ m_1; y(t))  \,,
\end{equation}
since it satisfies the conditions that $F_2(0) = 1$ and $F_2(y)$ be
analytic at $y=0$.

Using this solution as in the previous section, we can extract the
amplitude for the $2 \chi$ to $n \phi$ process by looking
at the terms containing both $z_0^n$ and the pole at $E = n m_1/2$.
Expanding $F_2(y)$ in powers of $y$,
\begin{equation}
    F_2 (y) = \sum_{k=0}^{\infty}
	{\Gamma (a + k) \over \Gamma (a)}
	{\Gamma (1 -a + k) \over \Gamma (1 - a)}
	{\Gamma (1 - 2 E / m_1) \over \Gamma (1 - 2 E / m_1 + k)}
	{y^k \over k!} \,,
\end{equation}
noting that the coefficient
\begin{equation}
    {\Gamma (1 - 2E/m_1) \over
	\Gamma (1 - 2E/m_1 + k)}
\end{equation}
in the $y^k$ term contains simple poles at $E = l m_1/2$ with
$l = 1, 2, \ldots , k$, and noting that $y^k$ has power expansion in $z_0$
starting from $z_0^k$, determines that only the term containing $y^n$
contributes to $A_n$.
Following steps similar to those in previous section to get the correct
normalization factor, we find that the amplitude is given by
\begin{equation}
    A_n = {\Gamma (a + n) \over \Gamma (a)}
	{\Gamma (1 - a + n) \over \Gamma (1 - a)} \,(n\, m_1)^2 \,
	{1 \over n!}
	\left ( 1 \over 2 \phi_0 \right )^n  \,.
\end{equation}
When the parameter $a$ takes non-integral values, the threshold
amplitude at tree-level for producing a large number of final particles,
$n \gg 1$, has the behavior
\begin{equation}
    A_{n} \approx n \,n! \, {m_1^2 \over \Gamma (a) \Gamma (1-a)}
	\left ({\lambda \over 12 m_1^2} \right )^{n/2} \,.
\end{equation}
However, when the parameter $a$ takes an integer value $N$, the
factor
\begin{equation}
    {\Gamma (1 -a + n) \over \Gamma (1 - a)} =
	{\Gamma (1 - N + n) \over \Gamma (1 - N)}
	= (1-N) (2-N) \cdots (n - N)
\end{equation}
vanishes for $n \ge N$, and thus the amplitude $A_n$ vanishes for
$n \ge N$.
In particular, for $a=3$, $A_n$ vanishes for $n > 2$,
which is the result of Smith \cite{Smith}.

\acknowledgments

We should like to thank P. Arnold, M. B. Voloshin, and L. G. Yaffe
for providing beneficial discussions.
The work was supported, in part, by the U. S. Department of Energy
under grant DE-AS06-88ER40423.

\end{document}